\documentstyle[preprint,aps]{revtex}
\begin{document}
\draft

\title{ LOCV Calculation for Beta-Stable Matter
at Finite Temperature}
\author{
 M. Modarres$^{a}$ and H. R. Moshfegh$^{b}$
}
\address{$^{a,b}$Physics Department, Tehran University, Tehran Iran \\
$^{a,b}$Institute for Studies in Theoretical Physics and Mathematics, Tehran,
Iran \\
$^{a}$Center for Theoretical Physics and Mathematics, P.O.Box 11365-8486,
 Tehran, Iran
}
\maketitle
\begin{abstract}
The method of lowest-order constrained variational, which predicts reasonably
the nuclear matter semi-empirical data is used to calculate the equation of
state of beta-stable matter at finite temperature. The Reid soft-core
with and without the N-$\Delta$ interactions which fits the N-N scattering
data as well as the $UV_{14}$ potential plus the three-nucleon interaction
 are considered in the nuclear many-body Hamiltonian. The electron and
muon are treated relativistically in the total Hamiltonian at given
 temperature, to make the fluid electrically neutral and stable against
 beta decay. The calculation is performed for a wide range of baryon density
and temperature which are of interest in the astrophysics. The free energy,
entropy, proton abundance, etc. of nuclear beta-stable matter are calculated.
 It is shown that by increasing the temperature, the maximum proton abundance
 is pushed to the lower density while the maximum itself increases as we
 increase the temperature. The proton fraction is not enough to see any
 gas-liquid phase transition. Finally we get an overall agreement with other
 many-body techniques, which are available only at zero temperature.
\end{abstract}
\section{Introduction}
The prediction of  equation of state of hot and dense nuclear matter
has been a subject of growing interest in nuclear physics and astrophysics.
Such matter under such unusual conditions, beside of various nuclear physics
applications, it is likely to be produced through stellar collapse,
supernova explosion, neutron stars etc. [1].

 On the other hand, it is believed
 that at extremely high density, deconfinement
can take place and the hadronic
matter may make a transition to the quark matter.
So the properties of transition points depend on the equation of state of
hadronic matter.

It is well known that the presence of leptons is also crucial since the
star matter have to be equilibrated against the weak leptonic-decay. So the
increase of charged leptons with certain baryonic density will finally
turns to a large fraction of neutron density. However, other negatively
charged baryons such as $\Sigma^-$ hypron may also become energetically
 more favorable [2].

  In order to study such systems one needs a good many-body techniques to
provide an accurate description of the equation of state, entropy and
other thermodynamics variables of hot nucleonic matter.

   In a series of papers the lowest order constrained variational (LOCV)
method has been developed [3-8] for calculating the properties of
homogeneous nuclear fluids with realistic nucleon-nucleon interaction. This
approach has been further generalized to include more sophisticated
interactions such as the $V_{14}$ [8], the $AV_{14}$ and the new
argonne $AV_{18}$ [8] as well as the Reid [3] and $\Delta$-Reid
[3] potentials. For a wide range of models our LOCV calculation agrees
well with the results of fermion hypernetted chain (FHNC) calculations
where these have performed and for a number of central potentials there is
agreement with the essentially exact numerical solutions obtained by Monte
Carlo technique [3]. Despite this agreement for model problems, there
has been some dispute about convergence of LOCV results in calculation
employing realistic nucleon-nucleon interactions which are strongly
spin-dependent and which, in particular, contain a sizeable tensor force.
This argument was tested by us by calculating the energy of the
three-body cluster
contribution in nuclear matter and the normalization integral $<\psi \mid \psi
>$ both at zero and finite temperature [4,7]. It was shown that
$<\psi \mid \psi>$ is normalized within one percent and the three-body
cluster energy is less than one MeV for $k_f \leq 1.6 fm^{-1}$. Our LOCV
calculation is a fully self-consistent technique and is capable of using
the well-defined phenomenological potentials such as 
 $\Delta$-Reid (the modified Reid potential with an
allowance of $\Delta (1234)$ degree of freedom, see Modarres and Irvine
[3]) potential. The $\Delta$ state is most important configuration
which modifies the nuclear force and it might be at the origin of
understanding of three-body forces [9]. The results suggest that
the LOCV method reasonably describes the nucleonic-matter properties at
zero and finite temperature.

On the other hand, our recent calculation at zero temperature [8]
with $V_{14}$ and $V_{18}$ potentials show the same behavior  and a very
good agreement was found with more sophisticated calculation such as the
fermion hypernetted chain method (FHNC) [10].

 With respect to the above arguments, in this work we shall attempt to
calculate
the properties of beta-stable matter at finite temperature by using the Reid
and $\Delta$-Reid potential and investigate the behavior of proton abundance
with the temperature and density of baryonic matter.

   Brueckner theory [11] and variational FHNC [12] have been also applied to
beta-stable matter but only for zero temperature. So we can also compare our
result with above calculations.

  The paper is planed in the following steps:
The beta-stable matter free energy is explained in section II. Section III
is devoted to a short description of the lowest order constrained variational
method. Finally in section IV we present the results and discussion.
\section{The beta-Stable matter free-energy}
The free energy of beta-stable matter is written as the sum of free energy
of the leptonic $(F_L)$ and the baryonic $(F_B)$ parts :
\begin{equation}
F=F_L+F_B
\end{equation}
The nucleons are assumed to interact through one of the realistic N-N
interactions i.e. Reid, $\Delta$-Reid and $V_{14}$ potentials. The requirement
of charge neutrality implies that we can ignore the electromagnetic interactions
and the week interactions are neglected.

 The total baryon number density $n_B$ is the sum of the proton and neutron
 number densities
\begin{equation}
n_B=n_p+n_n
\end{equation}
and the condition of electrical neutrality requires:
\begin{equation}
n_p=n_e+n_{\mu}
\end{equation}
(here we ignore $\tau$ leptons because of its large rest mass respect
to two other leptons.)

 The leptons form two highly relativistic fermi seas. The contribution to
 the energy per baryon from these fermi seas is,
\begin{equation}
E_L=(\Omega n_B)^{-1}\sum_{i=e,\mu}\sum_{k,\sigma}
[\varepsilon_i(k)+m_ic^2]f_i(k)
\end{equation}
where
\begin{equation}
\varepsilon_i(k)=[\hbar^2k^2c^2+m_i^2c^4]^{1/2}-m_ic^2
\end{equation}
and
\begin{equation}
f_i(k)=[\exp((\varepsilon_i(k)-\mu_i)\beta)+1]^{-1}
\end{equation}
is the familiar Fermi-Dirac distribution with $\beta={1 \over {K{\cal T}}}$
($K$ is the Boltzman factor). $\mu_i$ are the chemical potentials of the
$i$th pecies of particle.

 For zero temperature, $E_L$ can be calculated analytically and it take the
following form:
\begin{equation}
E_L(T=0)=\sum_{i=e,\mu}{{m_i^4c^5}\over{8\pi^2\hbar^3n_B}}[x_i(1+x_i^2)^{1/2}
(2x_i^2+1)-\sinh^{-1}x_i]
\end{equation}
where
\begin{equation}
x_i={{{\hbar}k_i}\over{m_ic}}
\end{equation}
and $k_e$ and $k_{\mu}$ are the electron and muon Fermi momenta respectively.
 The chemical potentials $\mu_i$ (or Fermi seas in case of ${\cal T}=0$) are
 related through condition of beta stability i.e.
\begin{equation}
\mu_n-\mu_p=\mu_e=\mu_\mu
\end{equation}
For ${\cal T}=0$ the second equality in above equation implies that
\begin{equation}
m_ec^2(1+x_e^2)^{1/2}=m_{\mu}c^2(1+x_{\mu}^2)^{1/2}
\end{equation}
while for ${\cal T}\neq0$ the chemical potentials are fixed by the various
particle number densities i.e.
\begin{equation}
n_i=(\Omega^{-1}) \sum_{\sigma,k} f_i
\end{equation}
and the second equality in equation (9) should be fullfiled numerically. It
is clear that as far as $\mu_e$ is less than the rest-mass of muon we will
not have any muon in the matter.

The first equality in equation (9) is satisfied by minimizing the total
free energy respect to the proton number density $n_p$.

  Then the leptonic free energy per baryon can be written as
\begin{equation}
F_L=E_L-{\cal T}S_L
\end{equation}
where
\begin{equation}
S_L=S_e+S_\mu
\end{equation}
and
\begin{equation}
S_i=K(\Omega n_B)^{-1}\sum_{k,\sigma}(1-f_i(k))
\ln (1-f_i(k))+f_i(k) \ln (f_i(k))
\end{equation}
Finally we write the baryonic free energy as
\begin{equation}
F_B=E_B-{\cal T}S_B
\end{equation}
where
\begin{equation}
S_B=S_p+S_n
\end{equation}
with definitions of $S_i$ from equation (14) and Fermi-Dirac distribution of
equation (6), but with $\varepsilon_i={{\hbar^2k_i^2}\over{2m_i^\ast}}$.
$m_i^\ast$ are the proton and neutron effective masses and they
will be treated
variationally. The baryonic internal energy is
\begin{equation}
E_B=T_B+E_B^{MB}
\end{equation}
where the kinetic energy part has the following form (the baryons are
treated non-relativistically),
\begin{eqnarray}
T_B=\sum_{i=n,p}{n_i\over n_B}({3\over 5}{{\hbar^2}\over{2m_i}}k_i^2
+m_ic^2)\hspace{2.79cm}if\hspace{1.5cm} {\cal T}=0\nonumber\\
\hspace{3.5cm}=(\Omega n_B)^{-1}\sum_{k,\sigma}\sum_{i=p,n}
[{{\hbar^2k^2}\over{2m_i}}+m_ic^2]
f_i(k)\hspace{1.5cm}if \hspace{1.5cm}T\neq 0
\end{eqnarray}
and $k_n$ and $k_p$ are the familiar neutron and proton fermi
momenta, respectively.
The many-body energy term $E_B^{MB}$ will be discussed in the next section.
\section{The LOCV formalism and $E_B^{MB}$ calculation}
The calculation follows exactly that of asymmetric nuclear matter
calculation of reference [6]. We consider an ideal Fermi gas type wave
function for the single particle states. Then using variational techniques,
we can write the wave function of interacting system as,
\begin{equation}
\psi={\cal F}_{\cal T}\Phi^{\cal T}
\end{equation}
where
\begin{equation}
{\cal F}_{\cal T}={\cal S} \prod_{i>j} f(ij)
\end{equation}
The Jastrow correlation functions $f(ij)$ are operators that act on spin,
isospin and relative coordinate variables of
particles $i$ and $j$ and $\cal S$ is
a symmetrizing operator which is necessary, since the $f(ij)$ do not
commute. The unitariness of ${\cal F}_{\cal T}$'s usually
cause the problem of
nonorthogonality of different states. But since at low temperature only
the one quasi particle-type states are important and these states have
different total momentum, they can be considered to be orthogonal.

   Now, using the above trial wave function, we can construct
a cluster expansion
for the expectation value of the following Hamiltonian,
\begin{equation}
H=\sum_i-{{\hbar^2}\over{2m}}\nabla_i^2 + \sum_{i<j}v_{ij}
\end{equation}
where $v_{ij}$ is a two-nucleon potential that fit the nucleon-nucleon
scattering data and deuteron properties. In this work we will mainly focus
on the Reid [3] and $\Delta$-Reid [3] interactions. Since, regarding our
previous works, the $\Delta$-Reid interaction can reasonably reproduce
the nuclear matter properties [3,8]. We will also use the $UV_{14}$ plus
density dependence three-body potential in order to compare our result
with others [12] at zero temperature.

In the cluster expansion series, we keep only the first two terms
of energy functional:
\begin{equation}
E_B([f])={1\over A} {{<\psi \mid H \mid \psi>}\over{<\psi \mid \psi>}}=
E_1+E_2
\end{equation}
$E_1$ is independent of $f(ij)$ and is simply the Fermi-gas kinetic energy
of baryons $T_B$ which already has been discussed in equation (18). While
the two-body energy $E_2$ is written as
\begin{equation}
E_2= {1\over {2A}} \sum_{ij}<ij \mid{\cal V}(12)\mid ij>_a
\end{equation}
where
\begin{equation}
{\cal V}(12)=-{{\hbar^2} \over {2m}}[f(12)\  ,\  [\nabla_{12}^2\   ,\  f(12)
]]+f(12)v(12)f(12)
\end{equation}
and the two-body antisymmetrized matrix element $<ij \mid {\cal V} \mid ij>_a$
is take with respect to the single-particle wave function $\Phi^{\cal T}$.

We minimize the two-body energy $E_2$ with respect to the variations in
functional $f(ij)$ but subject to normalization constraint [3,8],
\begin{equation}
{1 \over A}\sum_{ij}<ij \mid h^2(12)-f^2(12) \mid ij>_a=0
\end{equation}
where
\begin{eqnarray}
h=[1-{1 \over 2}({{\gamma_i(r)}\over {n_B}})^2]^{-{1\over 2}}
\hspace{1cm}  n-n \hspace{.25cm}and  \hspace{.25cm}p-p \hspace{.25cm} 
channel\nonumber\\
\hspace{1.5cm}=1\hspace{6.15cm} n-p \hspace{.25cm} channel
\end{eqnarray}
with ($i=p,n$)
\begin{equation}
\gamma_i(r)={4 \over {(2 \pi )^3}}\int f_i(k)J_0(kr) \, d\bf{k}
\end{equation}
The above constraint forces the cluster series to converge very rapidly and
we can approximate $E_2$ by $E_B^{MB}$ i.e. the many-body contribution to
the internal energy.

 The detail about the Euler-Lagrange differential equations, the boundary
conditions and the channel break down of $E_2$ have been discussed in
references [3,6].
\section{Results and Discussion}
The equilibrium configuration of beta-stable matter is obtained at
each total baryon number density $n_B$ and temperature ${\cal T}$ by
minimizing the total free energy $F$
 with respect to the two-body correlation functions, effective masses and
 proton abundance $n_p/n_B$ subject to the constraints of equations (2),(3),
 (9) and (25).

 In figure 1 we plot our calculated free energy per baryon number at
different temperatures for the Reid and $\Delta$-Reid potentials. This
figure can be compared with our previous calculation for pure neutron matter
[5] (figure 2). It is seen that beta-stable matter has softer equation
of state than pure neutron matter and as we expected the free energy
per baryon decreases with increasing temperature at constant density. But
with increasing the density the free energy shows an increasing trend for
given temperature. Inclusion of isobar, increases the free energy and
makes it more density dependent.

 The relative proton abundance, $y_p={{n_p}\over{n_B}}$, for the Reid and the
 $\Delta$-Reid potentials at different temperatures is plotted in figure 3. It
 is seen that $y_P$ increases at low baryon density as we increase the
 temperature and the maximum value of $y_P$  is pushed to the lower baryon
density by an increase in the temperature. This
shows that at high temperatures
the beta-stable matter with higher proton abundance tends to lower its baryon
density. This effect is much larger in case of the $\Delta$-Reid interaction.

The proton abundance peak is due to this fact that the system takes
the advantage of the spin-triplet tensor components in the Reid (or $\Delta$-Reid)
interaction and by maximizing the proton density at given
baryon density makes the free energy
lower. Since the $\Delta$-Reid interaction has more repulsive components,
 like $^1S_0\rightarrow \  ^1D_5$ transition, the peak
 is happened in lower baryon density.

 In figure 4 the proton, electron and muon number density are plotted
 against baryon number density at $K {\cal T}=5$ and $20$ MeV. We find
 that $n_e, n_{\mu}$ and $n_p$
increase by increasing temperature at low baryonic density
while their maximum values become lower as we
increase the temperature. The entropy
of proton, electron and muon are plotted versus baryonic density in
figures 5 and 6 for the Reid and $\Delta$-Reid potentials at
different temperatures.
It is seen that the entropy of proton and electron increase very rapidly
with increasing the temperature at low densities while for the
muon's entropy we find 
a peak for given temperature. The baryon density in which the muon's entropy becomes
maximum goes to lower densities as we increase the temperature.

In figure 7 we compare our beta-stable results for $y_p$ with those of
Wiringa et al. (WFF) [12] and Baldo et al. (BHF) [11]
for $UV_{14}$ and ${AV_{14}}$ potentials at zero
temperature. We get quite good agreement with them especially at
low baryonic densities. For $n_B \geq .5$ the BHF calculation increases while
our results and WFF decreases. 
Our results with Reid and $\Delta$-Reid potentials at zero temperature
are also given for comparison.

We would conclude by pointing out that we have carried out a calculation
of beta-stable matter using techniques, which we believe to be more
reliable than Brueckner calculation especially at higher densities
($n_B \geq 0.2 fm^{-3}$) and it is comparable with the variational
hypernetted chain method. It was seen that the extra degree of freedom offerd
by beta decay leads to a slight softening of the equation of state, as one
would expect. We found that, the maximum values of proton
abundance which are predicted by the Reid
and $\Delta$-Reid potentials are much smaller
than $UV_{14}$ and $AV_{14}$ potentials and they happened in much lower
baryonic densities. Increasing the temperature  of the beta-stable matter
increases the proton abundance in low densities. This effect becomes more
stronger when $\Delta$-Reid interaction is considered. In general, the proton
fraction is not enough to se the gas-liquid phase transition. It also dose
not reach the critical value of about 15 percent needed for the occurrence
of direct Urca Processes which believed to be responsible for a fast neutron
star cooling [1]. However it was shown that the temperature may play a
great role to reach to higher percentage of proton abundance and lower
baryonic density.

Finally the equation of state of beta-stable is more soften than neutron
matter and the effect of temperature may influence the behavior of dense matter
and the structure of neutron star.


\begin{figure}
\caption{
Free energy of beta-stable matter (without baryon rest masses)
(MeV) versus baryon density ($fm^{-3}$) at various temperatures (MeV).
Dashed curve and full curves are for the $\Delta$-Reid and Reid
potentials, respectively. }
\label{fig1}
\end{figure}
\begin{figure}
\caption{Same as figure 1 but for neutron matter.
}                  
\label{fig2}
\end{figure}
\begin{figure}
\caption{The proton abundance for the $\Delta$-Reid (dashed curve) and Reid
(full curve ) potentials at various temperatures (MeV).
}                                                  
\label{fig3}
\end{figure}
\begin{figure}
\caption{Particle densities ($fm^{-3}$) of $e, \mu$ and $p$ for the $\Delta$-Reid
(dashed curve) and Reid (full curve) potentials at $K{\cal T}=5$ and $20$
MeV.
}         
\label{fig4}
\end{figure}
\begin{figure}
\caption{The entropy of proton (Mev$^{-1}$) in the beta-stable matter for the
$\Delta$-Reid (dashed curve) and Reid
(full curve ) potentials at various temperatures (MeV).
}
\label{fig5}
\end{figure}
\begin{figure}
\caption{
As figure 5 but for $e, \mu$.
}
\label{fig6}
\end{figure}
\begin{figure}
\caption{The comparison of proton abundance $y_p$ versus baryon density
($fm^{-3}$) at zero temperature  with those of references [11,12] for
te $UV_{14}, AV_{14}$, Reid and $\Delta$-Reid potentials. 
}
\label{fig7}
\end{figure}


\begin{thebibliography}{99}
\bibitem[1]{sto} H. Stocker and W. Greiner, Phys. Rep. 137 (1986) 137.\\
I. Bombaci, T. T. S. Kuo and U. Lombardo, Phys. Rep. 242 (1994) 165.\\
M. Prakash et al. Nucl-th /9603042.\\
J. Lattimer, C. Pethick, M. Prakash and P. Haensel, Phys. Rev. Lett 66 (1991) 2701.
\bibitem[2]{b} M. Baldo, G. F. Burgio and H. G. Schulze, Phys. Rev. C58 (1998) 3688.\\
I. Vidana, A. Polls and A. Ramos, nucl-th/9909019
\bibitem[3]{pra} J. C. Owen, R. F. Bishop and J. M. Irvine, Ann. Phys. N. Y.
102 (1976) 170.\\
M. Modarres and J. M. Irvine, J. Phys. G: Nucl. Phys. 5 (1979) 511.
\bibitem[4]{latt}  M. Modarres and J. M. Irvine, J. Phys. G: Nucl. Phys. 5 (1979) 7.
\bibitem[5]{su}M. Modarres, J. Phys. G: Nucl. part. Phys. 19 (1993) 1349. \\
M. Modarres, J. Phys. G: Nucl. part. Phys. 21 (1995) 351.
\bibitem[6]{su}M. Modarres, J. Phys. G: Nucl. part. Phys. 23 (1997) 923.
\bibitem[7]{m} H. R. Moshfegh and M. Modarres, J. Phys. G: Nucl. part. Phys. 24 (1998) 
821.
\bibitem[8]{g} G. H. Bordbar and M. Modarres,  J. Phys G:Nucl. Part. Phys. 23 (1997) 
1631. \\
 G. H. Bordbar and M. Modarres,  Phys. Rev. C57 (1998) 7114.\\
 M. Modarres and G. H. Bordbar,  Phys. Rev. C58 (1998) 2781.
\bibitem[9]{Heyd} K. Heyde "Basic Ideas and Concepts 
in Nuclear Physics" IOP publishing, Bristol (1994).
\bibitem[10]{friedman} B. Friedman and V. R. Pandharipande, Nucl. Phys.
A361 (1981) 502.
\\K. E. Schmidt and V. R. Pandharipande, Phys. Lett. B87 (1979) 11.\\
I. E. Lagaris and V. R. Pandharipande, Nucl. Phys. A359 (1981) 331.\\
A. Akmal, V. R. Pandharipande and D. G. Ravenhall, Phys. Rev. C58 (1998) 1804.
\bibitem[11]{mm}M. Baldo, G. F. Burgio and H. J. Chulze, Phys. Rev. C58 (1998) 3688.
\\M. Baldo, I. Bombaci and G. F. Burgio, Astron. Astrophys. 328 (1997) 274.
\bibitem[12]{w}R. B. Wiringa, V. Fieks and A. Fabrocini, Phys. Rev. C38 (1988) 1010.
\end{thebibliography}
\end{document}